%% file: output.tex
\documentclass[conference]{IEEEtran}
\IEEEoverridecommandlockouts
\usepackage{cite}
\usepackage{amsmath,amssymb,amsfonts}
\usepackage{algorithmic}
\usepackage{graphicx}
\usepackage{textcomp}
\usepackage{comment}
\usepackage{caption}
\usepackage[linesnumbered,ruled,vlined]{algorithm2e}
\usepackage{listings}
\usepackage{verbatim}
\usepackage{amsmath}
\usepackage{adjustbox}
\usepackage[table,xcdraw]{xcolor}
\usepackage{tikz}
\usepackage[normalem]{ulem}
\usepackage{url}
\DeclareMathOperator*{\argmax}{argmax}

\def\BibTeX{{\rm B\kern-.05em{\sc i\kern-.025em b}\kern-.08em
    T\kern-.1667em\lower.7ex\hbox{E}\kern-.125emX}}
\begin{document}

\title{ExplFrame: Exploiting Page Frame Cache for Fault Analysis of Block Ciphers}

 \author{\IEEEauthorblockN{Anirban Chakraborty}
 \IEEEauthorblockA{\textit{Indian Institute of Technology}\\
 \textit{Kharagpur, India}\\
 anirban.chakraborty@iitkgp.ac.in}
 \and
 \IEEEauthorblockN{Sarani Bhattacharya}
 \IEEEauthorblockA{\textit{Indian Institute of Technology}\\
 \textit{Kharagpur, India}\\
 tinni1989@gmail.com}
 \and
 \IEEEauthorblockN{Sayandeep Saha}
 \IEEEauthorblockA{\textit{Indian Institute of Technology}\\
 \textit{Kharagpur, India}\\
 sahasayandeep@cse.iitkgp.ac.in}
 \and
 \IEEEauthorblockN{Debdeep Mukhopadhyay}
 \IEEEauthorblockA{\textit{Indian Institute of Technology}\\
 \textit{Kharagpur, India}\\
 debdeep@cse.iitkgp.ac.in}
}

\maketitle

\begin{abstract}

\emph{Page Frame Cache (PFC)} is a purely software cache, present in modern Linux based operating systems (OS), which stores the page frames that are recently being released by the processes running on a particular CPU. In this paper, \emph{we show that the page frame cache can be maliciously exploited by an adversary to steer the pages of a victim process to some pre-decided attacker-chosen locations in the memory}. We practically demonstrate an end-to-end attack, \emph{ExplFrame}, where an attacker having only user-level privilege is able to force a victim process's memory pages to vulnerable locations in DRAM and deterministically conduct Rowhammer to induce faults. We further show that these faults can be exploited for extracting the secret key of table-based block cipher implementations. As a case study, we perform a full-key recovery on OpenSSL AES by Rowhammer-induced single bit faults in the T-tables. We propose an improvised fault analysis technique which can exploit any Rowhammer-induced bit-flips in the AES T-tables.
To the best of our knowledge, this is the first work highlighting the vulnerabilities of PFC and fault analysis of block cipher using Rowhammer completely from user-space.
\end{abstract}


\input{source/introduction.tex}
\input{source/background.tex}

\input{source/attack.tex}

\input{source/evaluation.tex}
\input{source/analysis.tex}
\input{source/conclusion.tex}

\bibliographystyle{./bibliography/IEEEtran}
\bibliography{./bibliography/IEEEabrv,./bibliography/IEEEexample}

\end{document}

%% file: source/introduction.tex
\section{Introduction}
\label{sec:intro}

Modern operating systems (OS) are optimized to obtain the best possible performance and throughput on a given hardware architecture.
The memory allocation mechanism of OS plays a crucial role in determining the overall performance of a system. The efficiency of this OS subsystem is mainly attributed to its intelligent usage of \emph{caching}, which helps in taking advantage of the locality of reference (temporal and spacial) in the memory hierarchy. 

Memory allocation subsystems in modern Linux-based OS use \emph{Buddy Allocation scheme} to allocate memory pages to different processes. When a process requests for memory, the buddy allocator allocates the required amount of memory in the form of fixed sized \emph{page frames}. As the process terminates, the allocated page frames are added back to the memory pool. To boost memory performance, the kernel maintains a per-CPU \emph{page frame cache} (PFC) which is a small software cache storing recently de-allocated page frames. Upon the arrival of a new request, the page frames inside the PFC are the first to serve it, before going to the actual memory pool. Moreover, this allocation scheme is oblivious to the processes and, in practice, the pages left by one process (in PFC) can readily be re-allocated to another process.     

The aim of this work is to maliciously exploit the aforementioned, seemingly benign, memory caching policy defined in the buddy allocator. We exploit the fact that the allocation of pages from PFC does not take the identity of the processes into account. Using this property of the PFC, we show that an adversary, having only user privilege, can indirectly steer the pages of a victim process to some pre-determined memory locations inside the  Dynamic Random Access Memory (DRAM). Restricting a victim process to operate in some attacker-controlled memory locations may have severe security implications. Here we show that even mathematically robust cryptosystems can fall prey to this vulnerability.  

One of the most prominent DRAM vulnerabilities known till date is the \emph{Rowhammer bug}~\cite{kim_isca14}. It is a phenomenon observed in most of the modern commercial DRAM modules where repeated access to a particular row induces bit flips in one of the adjacent rows. However, inducing precise faults using Rowhammer is a challenge due to the  uncontrollability of flip locations which is specific to a DRAM instance. In practice, some of the rows might show higher chances of getting faulted than others.
However, if the pages of a victim process get assigned to a Rowhammer vulnerable location, faults can be induced in the process in a regular manner. Quite obviously, PFC becomes a nice tool in this context as it can force the pages of a victim to some attacker-decided memory regions. Putting it differently, Rowhammer provides a concrete use case for showing the exploitability of the PFC allocation scheme. In this paper we present an end-to-end practical realization of the aforementioned idea of combining PFC with Rowhammer. The proposed attack strategy, called \textbf{\textit{ExplFrame}}, has been utilized to launch a practical key recovery attack on the T-table-based AES implementation from OpenSSL 1.1.1~\cite{openssl}. 

Previous work in \cite{vanderVeen_drammer16, kwong_rambleed, vanveen_guardion} have exploited Linux's Buddy Allocation system to perform \emph{memory massaging} for conducting Rowhammer. In particular, they exhaust the memory pages during templating phase, which is dependent on the allocation policy of the Buddy Allocator. 
\emph{Our proposed ExplFrame does not rely on the allocation policy; rather we exploit the principle of caching in one of the components of memory allocation subsystem, the PFC. Most importantly, the exploitation of PFC, to the best of our knowledge, has never been used as an attack vector before.} 
It is worth mentioning that table-based AES implementations have previously been targeted with Rowhammer-induced faults in~\cite{zhang_pfa18}. The main idea is to corrupt an entry inside the T-Table and thereby create a statistical bias within AES state, which can be used for key recovery. However, one immediate advantage of doing the fault attack with ExplFrame strategy is that it can be done from user privilege level. In contrast, the attack proposed in~\cite{zhang_pfa18} uses \texttt{pagemap} which requires administrative privilege on modern Linux distributions. Moreover, we observed that the strategy of fault exploitation in~\cite{zhang_pfa18} (called Persistent Fault Attack or PFA), is limited by the fact that it can only recover the key if certain specific bits of the T-table get affected by fault injection~\footnote{The reason for this will be explained later in this paper.} 
This being practically infeasible, the original PFA proposal cannot be used in the present context. Hence, as a second contribution, we propose a general fault exploitation methodology called Deep Round Persistent Fault Attack (DRPFA), which can exploit any Rowhammer-induced bit flip within the T-table. Further, we comment that Error Correcting Codes (ECC), which are often considered as effective countermeasures against Rowhammer induced faults~\cite{kim_isca14}, are not sufficient for preventing exploitable information leakage. We present a brief discussion on this at the end of this paper.  %

The rest of the paper is organized as follows. We present a brief background on the memory allocation schemes and Rowhammer in Sec.~\ref{sec:background}, followed by an overview of our attack ExplFrame in Sec.~\ref{sec:attack}. In Sec.~\ref{sec:attacking_aes}, we demonstrate an end-to-end fault induction method on the T-tables of OpenSSL AES using ExplFrame. We further provide an improved key recovery algorithm with the induced faults in Sec.~\ref{sec:fault_attack} and experimental results in Sec.~\ref{sec:results}~\footnote{We have informed and shared our findings with Intel Product Security Incident Response Team.}. The applicability of the attack at different scenarios has been discussed in Sec.~\ref{sec: discussion}. Finally, we conclude in Sec.~\ref{sec:conclusion}.

%

%% file: source/background.tex
\section{Background}
\label{sec:background}

\subsection{Linux memory allocation subsystem}
In NUMA (Non-Uniform Memory Access) based OS, each \textit{node} \footnote{NUMA systems classifies memory into nodes. Each node has similar access characteristics and affinity to one processor.} is divided into a number of blocks called \textit{zone}. 
Inside each memory zone, the allocation process is handled by the core allocator for Linux, called \textit{Buddy allocator}. In this allocation scheme, the pages are clustered into large blocks of size in power of two. When a request for certain amount of memory comes from the processor, the algorithm first searches the blocks of pages to check if the request can be met. If no blocks of pages are found to meet the demand, block of the next size is split into half and one half is allocated to the requesting process. The two smaller blocks thus produced are called buddies to each other. The process of splitting a block into half continues until a block of desired size is obtained. 
Likewise, when the allocated block becomes free, the buddy block is also examined. If both the blocks are free, they are merged together and returned back to their original block size. The coalescing of blocks on de-allocation gives rise to the name of the allocation scheme. 


The OS maintains a \textit{page frame cache}~\footnote{Not to be confused with \textit{page cache} which contains files read from the disk, memory-mapped files, shared libraries, etc.} for each memory zone. This small software cache of recently de-allocated (released) page frames are used by the Buddy allocator if the local CPU requests a small amount of memory, typically a few pages. The presence of page frame cache can significantly boost up the system performance by taking advantage of the locality of reference. The OS kernel keeps track of two watermarks to monitor the size of the cache. Whenever the number of page frames in the cache falls below the low watermark, the kernel brings in more page frames from the buddy system. Similarly, when the number of page frame surpasses the high watermark due to release of page frames from different running and finished processes, the kernel releases some of the page frames back to the buddy system. 
\emph{It is worth mentioning that our attack exploits the caching mechanism of the allocator and is independent of the allocation policy itself.}

\subsection{The Rowhammer bug}
DRAMs have been constantly scaled down to accommodate larger number of memory cells into smaller physical space, thereby reducing the cost-per-bit of memory. However, cramming a large number of DRAM cells in small space leads to electromagnetic coupling effects among themselves. 
Owing to its closely packed architecture, when a particular DRAM row is accessed consistently and in high frequency, the cells in the neighbouring rows tend to lose their charge, thereby inducing bit-flips. This phenomenon is termed as Rowhammer bug, which has been exploited to launch several devastating classes of attacks in recent past \cite{vanderVeen_drammer16, kwong_rambleed, vanveen_guardion}.

The driving force behind Rowhammer bug is that specific DRAM rows must be repeatedly activated fast enough such that the adjacent rows lose charge. However to achieve this, the content of the cache memory must be flushed after every access so that all the requests are served from the main memory. In x86 architecture, flushing of cache can be achieved from userspace by the \texttt{clflush} command. This fact indicates that Rowhammer on standard x86-64 machines does not require any special privilege for repeatedly activating DRAM rows. However, in practice, only certain specific regions in a DRAM chip are found to be vulnerable under Rowhammer, and it is purely driven by the device physics. In order to practically induce faults, the target data must be located on a Rowhammer-vulnerable location in DRAM~\footnote{Once the vulnerable locations are identified, inducing bit-flips in them is repeatable.}. Hence, from an attacker's perspective, exploiting the Rowhammer bug is not trivial and creating a deterministic exploit requires certain other vulnerabilities to be combined with this one.

%% file: source/attack.tex
\section{The ExplFrame Approach}
\label{sec:attack}
Continuing our discussion from Linux memory subsystem, in this section we introduce the security implication of such performance improvisation scheme in details and propose ExplFrame, which utilizes PFC for performing Rowhammer almost deterministically on a victim process. 
\subsection{Exploiting Page Frame Cache}
Page Frame Cache stores the recently freed pages from all processes running on a particular processor core.
 The primary intention of PFC is to boost the performance of the memory allocation subsystem by keeping recently used page frames close to the processor, in case the process requests for additional memory in near future. However, the security implications of this simple scheme has never been analyzed in literature, and we explore in this work for the first time.
%
%
\vspace{-0.27cm}
\subsection*{\textbf{How can one exploit the PFC?}}
Let us consider the following example:
\begin{itemize}
    \item Process A is running on a standard system and requests some amount of memory (say by using \texttt{mmap} with the \texttt{MAP\_POPULATE} flag). On such instance, the buddy allocator scans the list of available page frame blocks and finds out a block that can satisfy the request.
    \item In course of time, the process A de-allocates a page (say by using the function \texttt{unmmap}). The `unmapped' page resides in the page frame cache of the zone in the anticipation that it could be requested by the same process A in recent future.
\end{itemize}
Thus if the same program requests for additional memory during the course of its execution, the OS attempts to serve the page frames from the cache. If the request is small, then it is satisfied by the cache; else, it will invoke the buddy allocator once more. The situation becomes interesting to a security engineer when there is another process (Process B) running simultaneously on the system and sharing the same CPU. Consider, process A is an adversary and process B is oblivious of such adversary.
\begin{itemize}
    \item Once again we have Process A running on the system which allocates some memory, unmaps one or two pages and waits\footnote{The adversarial process (Process A) must remain active rather than going into inactive state (sleep), since in that case the entire process state information including page frame cache will be swapped out of memory.}.
    \item In this scenario, Process B sends a request for additional memory pages. The OS will first try to service the request from the page frame cache itself. Thus, there is a high probability that the page frame that was unmapped by the adversarial process gets allocated to the victim.
\end{itemize}

Therefore, PFC can be exploited by an adversary to control the memory allocation of another process. More precisely, an adversary can restrict the physical memory locations that a legitimate process can use. As a practical implementation of our claim, we present ExplFrame, which uses PFC to steer a victim process's sensitive data into Rowhammer-vulnerable locations and subsequently induce faults using Rowhammer. 

 \subsection{Threat Model}
\label{sec:threat_model}
 We assume a multi-user server environment running on a Linux-based OS, where the adversary could introduce precise faults on victim's sensitive data. The adversary has user-level privileges. It is further assumed that the adversary cannot access any security sensitive memory possessed by the victim. This exploit requires that the victim and the adversary are operating on the same processor core~\footnote{Previous attack in \cite{kwong_rambleed} also assumes that the attacker and the victim are operating on the same core and binds the processes to the same CPU core using \texttt{taskset}.}.
 %
\subsection{Outline of the attack}
The adversary performs the following steps in order
\begin{itemize}
    \item She performs \emph{bin-partitioning} to partition the allocated memory space into bins to identify the individual DRAM banks. After the partitioning is complete, she conducts Rowhammer on one of the bins to find out a vulnerable page. We provide a detailed representation of bin-partitioning process in the next section.
    \item She unmaps the vulnerable page and due to the presence of PFC, the unmapped page gets cached in it. As discussed in previous subsection, the unmapped page stays in the PFC until some process requests for memory.
    \item She waits until a victim process requests for some memory. Due to the property of PFC, the pages in the cache are the first one to get allocated to the victim. Once the vulnerable page is allocated to the victim process, the adversary starts rowhammering once again in the same bin (which also contains the vulnerable page). 
\end{itemize}
We present a detailed description of all the aforementioned steps in a practical setting in the next section.

%% file: source/evaluation.tex
\section{Attacking AES T-tables: A case study}
\label{sec:attacking_aes}
ExplFrame is a generic attack which exploits the vulnerabilities of PFC to deterministically conduct Rowhammer on victim's data. In this section we present a practical end-to-end attack on OpenSSL AES using ExplFrame to induce faults in T-tables. In the next subsection, we present a novel memory partitioning algorithm in order to utilize rowhammer in a nearly deterministic way. 

\subsection{Deterministic Rowhammer from user-space}
\label{sec:overview}
One of the major challenges for precise Rowhammering on a particular location is that the physical layout of memory is abstracted by the OS. Also, modern Linux kernel does not allow access to \texttt{pagemap} from user privilege. Therefore, in order to perform Rowhammer from userspace on a specific memory page, we need to determine the DRAM bank where the page is located and repeatedly access (hammer) the addresses on that particular bank. 

\begin{figure}[h]
		\centering
		\includegraphics[height=4cm,width=\linewidth]{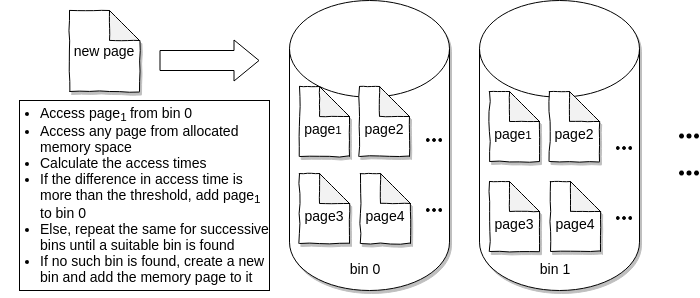}
		\caption{An overview of the bin partitioning process}
		\label{fig:bin_partitioning}
\end{figure}

Previous works \cite{pessl_drama} have shown that when a pair of addresses are accessed simultaneously, it creates a measurable timing channel depending on whether the address belong to different banks or same bank but different rows within the DRAM \footnote{If the addresses belong to the same bank but different row, then it will create a \emph{row conflict}. The first access will bring the data into the row buffer while at the time of second access, the first row will be closed first and then the second one is fetched. Due to row conflict, the difference in access time will be much higher than the other cases.}. Based on the DRAM access timing side-channel, we present a novel \textit{bin partitioning} technique to partition the entire allocated address space into the hypothetical `bins' such that each bin corresponds to a DRAM bank. 

An overview of the bin partitioning process is shown in Figure \ref{fig:bin_partitioning}. We first access the first page and put it in $bin_0$. Next, we access the next page and the first page simultaneously and check their access times. If the access time for the second page is more than some pre-defined threshold (can be determined by initial profiling of the system), that would mean the pair of accesses have resulted in a row conflict. So, the pair of page frames must be located in the same bank but different row. In that case, we put the second page in the same bin as the first one, i.e, $bin_0$. Whereas, if the access time is less than the threshold, we put it in the next bin, i.e, $bin_1$. Similarly, we pick the next page and check its access time with respect to the pages already stored in the bins. Based on the timing value, we put the page into one of the bins. This process is repeated until all the allocated pages are exhausted. In the end, all the page frames will be partitioned into separate bins.

\subsection{Putting it all together}
  We allocate a large memory (1 GB in our case) and partition the entire available memory into $n$ bins (n = 16 for the DRAM that we targeted) using \emph{bin partitioning} method, where each bin corresponds to a DRAM bank. After partitioning of the memory space, we start conducting Rowhammer by randomly picking addresses stored in last bin \footnote{Statistically, the last bin will have the smallest number of mismatches after two pass of the algorithm.}. The hammering is done for a stipulated time (1 hour in our case) and if no fault is found then we move on to the preceding bin. This process continues until a flip is found. If no such flip is found in any bin, the entire process is killed and restarted once again. If a flip is found, the corresponding page is unmapped.
\begin{figure}[h]
		\centering
		\includegraphics[height=5.5cm, width=\linewidth]{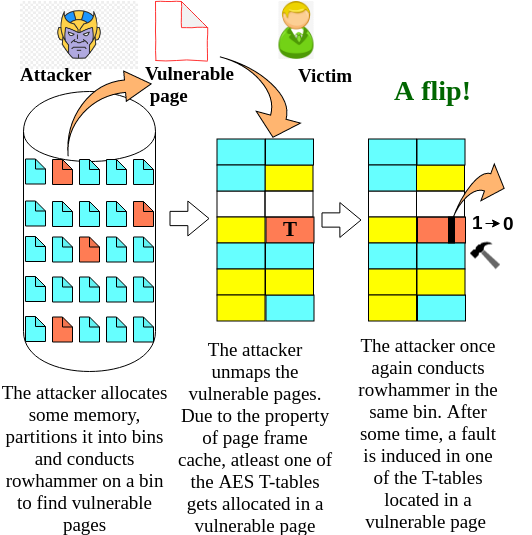}
		\caption{An overview of ExplFrame on OpenSSL AES T-tables}
		\label{fig:attack_overview}
\end{figure}

  We target the encryption T-tables ($T0$ through $T3$) of AES of OpenSSL 1.1.1 in a standard multi-user environment. Now, the adversary waits for the victim to load the T-tables into the memory. Due to the presence of PFC in each memory zone, the T-tables will be allocated in the same vulnerable page which was unmapped earlier. Once one of the T-tables is placed in a freed vulnerable page, the adversary again starts Rowhammering on the same bin. Due to reproducibility of the fault, the same page frame which now contains the T-table gets faulted once again, thereby corrupting the particular entry of the T-table. We validated our experiments on two standard Desktop computers having the specifications as mentioned in Table~\ref{tab:setup}. A pictorial representation of our ExplFrame attack on AES T-tables is depicted in Figure \ref{fig:attack_overview}.
\begin{table}[t]
    \centering
    \adjustbox{max width=\linewidth}{
    \begin{tabular}{|c|c|c|c|c|}
    \hline
    \rowcolor[HTML]{D6D6D6} 
    \textbf{Processor} & \textbf{\begin{tabular}[c]{@{}c@{}}Micro-\\ architecture\end{tabular}} & \textbf{\begin{tabular}[c]{@{}c@{}}DRAM type\\ \& capacity\end{tabular}} & \textbf{\begin{tabular}[c]{@{}c@{}}Operating\\ System\end{tabular}} & \textbf{\begin{tabular}[c]{@{}c@{}}Kernel\\ version\end{tabular}} \\ \hline
    \begin{tabular}[c]{@{}c@{}}Intel i5\\ 3330\end{tabular} & IvyBridge & \begin{tabular}[c]{@{}c@{}}Hynix DDR3 \\ 4GB\end{tabular} & Ubuntu 14.04 & \begin{tabular}[c]{@{}c@{}}4.13.0-36\\ generic\end{tabular} \\ \hline
    \begin{tabular}[c]{@{}c@{}}Intel i7\\ 7700\end{tabular} & KabyLake & \begin{tabular}[c]{@{}c@{}}Micron DDR4\\ 8GB\end{tabular} & Ubuntu 18.04 & \begin{tabular}[c]{@{}c@{}}4.15.0-50\\ generic\end{tabular} \\ \hline
    \end{tabular}
    }
    \caption{Experimental Setup}
    \label{tab:setup}
    \vspace{-0.6cm}
\end{table}
%



%% file: source/analysis.tex
\section{Practical Exploitation of Induced Faults}\label{sec:fault_attack}
In this section, we present a generic and practical methodology for exploiting the faults induced by Rowhammering. Given that the use-case of ours induces faults in AES T-tables, we aim to perform a key recovery attack on AES by analyzing these faults. There exist a large body of work addressing fault attacks (FA) on different classes of cryptographic primitives, especially on block ciphers like AES~\cite{aes_fault}. However, the attack algorithms vary largely depending on the type of the faults that can be practically induced within a system. One should note that the corruption in the present case happens in the T-tables of the implementation. The injected fault persists in the T-table until it is reloaded. This typical fault model matches with the one proposed by Zhang et. al. in~\cite{zhang_pfa18} (popularly referred to as Persistent Fault Model (PFM)). The main idea of PFA is to exploit the statistical bias resulting from the AES computation with a corrupted T-table. The faulty outcomes (ciphertexts) are analyzed statistically by guessing the last round key candidates. The statistical bias becomes visible only for the correct key guess eventually returning the key.  
The original proposal in~\cite{zhang_pfa18} makes two important assumptions:
\begin{enumerate}
    \item The adversary exactly knows the value that got corrupted inside the T-table.
    \item The corrupted entry in T-table is accessed at the last round of AES computation. This usually happens with a reasonable probability. 
\end{enumerate}

Several issues may arise while realizing the PFA from a practical perspective using Rowhammer. Below we enlist the main realization issues:
\begin{enumerate}
    \item \textbf{The Target Implementation:} The implementation under attack plays a crucial role for the success of PFA with Rowhammer induced faults.
    T-table-based implementations are available at state-of-the-art crypto-cores like OpenSSL, Libgcrypt etc. However, intricate implementation differences may still be there. As an example, we consider two competing implementations from OpenSSL and Libgcrypt. The first one utilizes 4 T-tables $T_0$, $T_1$, $T_2$ and $T_3$ for all the rounds $r_i$ ($1 \leq r_i \leq 10$). In contrast, Libgcrypt utilizes 4 alternative T-tables $T^{'}_0$, $T^{'}_1$, $T^{'}_2$ and $T^{'}_3$ to realize the last round of AES.
    \item \textbf{Nature of Rowhammer-induced Faults:} Rowhammer fault injection strategy provides limited control over the faults to be induced. More specifically, \emph{it is hard to control which bit of the T-table gets corrupted.}
\end{enumerate}

\begin{figure}[h]
    \centering
    \vspace{-0.2cm}
    \includegraphics[scale=0.35]{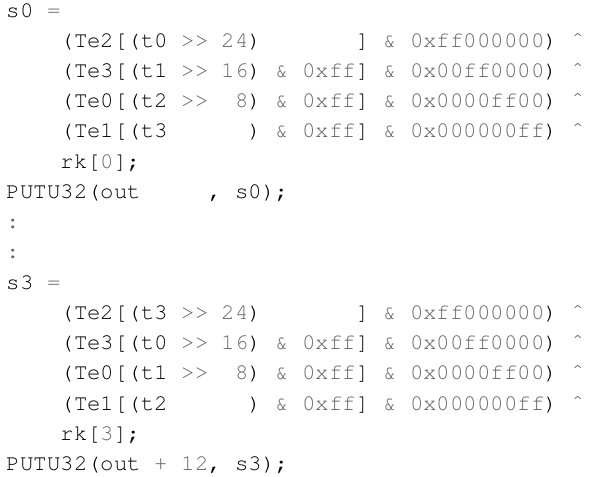}
    \caption{Code Snippet for AES Last Round in OpenSSL.}
    \label{fig:listing}
    \vspace{-0.15cm}
\end{figure}

{In~\cite{zhang_pfa18}, Zhang et. al. performed a case study on Libgcrypt implementation which uses separate T-tables for the last round. As a result, faults induced in their experiments only affect the last round.} As a more generic scenario, we consider the OpenSSL implementation for the illustration which uses the same T-tables for realizing all the AES rounds. Some part of the high-level C-code for the last round in OpenSSL is depicted in Fig.~\ref{fig:listing}. One should observe that \emph{in order to undo the effect of the AES MixColumns sub-operation, some of the bytes in a T-table output are masked (i.e. ANDed with zero).}
%
%

We next point out why these two above-mentioned realization issues are important from a practical perspective. Our first observation is that \emph{the original PFA algorithm does not work in this context with Rowhammer induced faults on OpenSSL AES implementation.} To elaborate this observation, we point out that \emph{three out of four bytes in each T-table output are masked in the final round computation of AES (in order to undo the effect of the MixColumns operation)}. {Now if the Rowhammer fault affects any of these masked byte locations, the fault effect will also get masked and would not propagate to the output. As the attacker cannot precisely control which bit of the T-table gets corrupted, the PFA attack targeting the last round is suppose to fail with a reasonably high probability.} In nutshell, the Rowhammer induced faults cannot be utilized with the PFA algorithm described in~\cite{zhang_pfa18} for many of the practical implementations like the one in OpenSSL.   

Fortunately, our investigation revealed that the faults induced by Rowhammering are still exploitable for key extraction. In this context we propose a novel and generic attack strategy called Deep Round Persistent Fault Attack (DRPFA) which is not affected by the fact that the attacker may not have precise control over the location of the Rowhammer induced faults. One should recall that the fault in the T-table is persistent, and as a result it also affects certain intermediate rounds of the AES computation. The proposed DRPFA attack exploits these corruptions in intermediate rounds for extracting the key. The advantage of this strategy is that we need not care about the precise location of the faulty bit anymore. The details of the DRPFA attack is presented in the following subsection.

\subsection{Deep Round Persistent Fault Attack}
Let us now describe the DRPFA attack algorithm. The pseudo-code for the algorithm is depicted in Algorithm~\ref{algo:drpfa}. The main idea is to guess a part of the secret and partially decrypt the ciphertexts up to the round where the statistical bias is being observed. More precisely, the partial decryption should continue up to the inverse MixColumns of the target round so that the bias at the S-Box outputs can be exploited. According to the well-known wrong-key assumption, the aforementioned statistical bias becomes visible for the correct key guess with a very high probability, and for the wrong guesses with negligibly small probability. This fact enables the recovery of the key with a fairly simple statistical test. We utilize the Squared-Euclidean-Imbalance (SEI) test for identifying the bias and thus the correct key. Also, in order to reduce the complexity, our attack mainly targets the 9th round of the AES computation. One important observation regarding DRPFA is that, the attack remains equally applicable even in the presence of combined Side-Channel Analysis (SCA) and Fault Attack (FA) countermeasures. From this perspective, DRPFA is equally powerful as of SIFA~\cite{dobraunig_sifa} and PFA~\cite{pan_onefault}. 

%
%

%
\begin{algorithm}[h]
    \scriptsize
    \DontPrintSemicolon
    \SetAlgoLined
    \KwIn{Ciphertexts from Rowhammer fault injection campaign ($\mathcal{C}$), target round $r$.}
    \KwOut{Key $k_c$}
        \textbf{set} $s :=$ size of the partial key guess based on $r$\; 
        \textbf{set} $K_g := \{0, 1, \cdot \cdot \cdot, 2^{s} - 1\}$\;
        \textbf{set} $SEI_{dict} := \emptyset$\;
        \For {$k_g \in K_g$ }
        {
          \textbf{set} $S_{k_g} := \emptyset$\;
          \For {$c \in \mathcal{C}$ }
          {
            $S_{k_g} := S_{k_g} \cup$ \texttt{Partial\_Decrypt}($c$, $r$)\;
          }
          $SEI_{dict}[k_g] := $\texttt{SEI}($S_{k_g}$)\;
        }
        \textbf{set} $k_c := \underset{k_g \in K_g}\argmax~SEI_{dict}[k_g]$\;
        \textbf{return} $k_c$\;
     \caption{The DRPFA Algorithm}
     \label{algo:drpfa}
\end{algorithm}

\vspace{-0.15cm}

\section{Experimental Validation}
\label{sec:results}

\begin{figure}[h]
    \centering
    \vspace{-0.2cm}
    \includegraphics[scale=0.30]{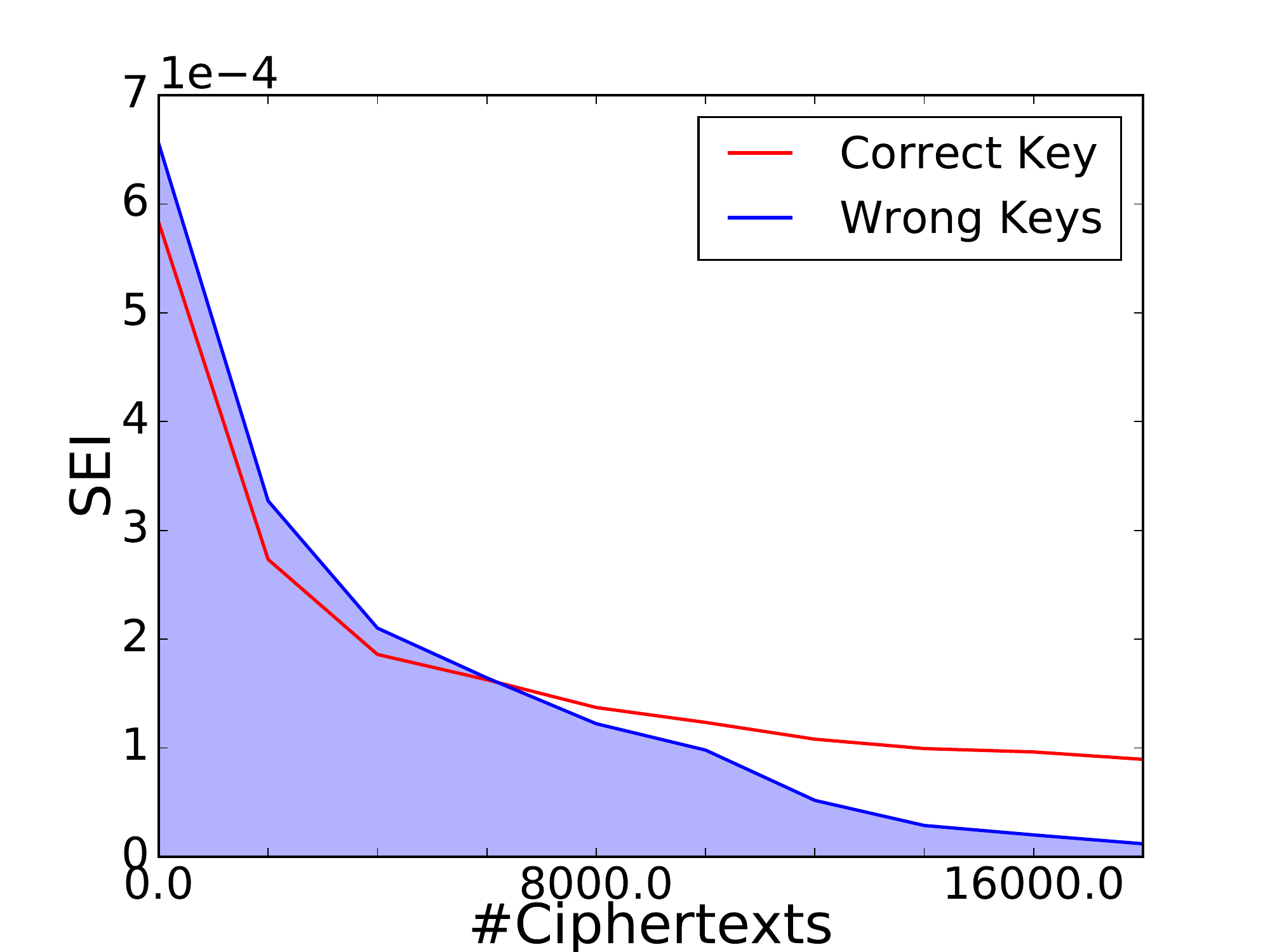}
    \caption{SEI convergence plot for the DRPFA attack.}
    \label{fig:plot_sei_convergence}
    \vspace{-0.15cm}
\end{figure}

The results of inducing faults in OpenSSL AES T-tables is shown in Table \ref{tab:results}. In order to validate the practicality of DRPFA with these faults, we encrypt $20000$ plaintexts after the fault is induced. Here the fault has been induced in the table T0, and is present throughout the encryption campaign. The 9th round S-Box output is considered for attack. Consequently, the key is extracted in chunks of $32$-bits (i.e. $s = 32$ according to Algorithm.~\ref{algo:drpfa}. It is worth mentioning that the PFA attack did not work with this specific fault even with $100000$ ciphertexts. However, DRPFA successfully extracts the key with roughly $8000$ ciphertexts. 
\begin{table}[h]
\centering
\adjustbox{max width=\textwidth}{
\begin{tabular}{|c|c|c|c|c|}
\hline
\rowcolor[HTML]{D6D6D6} 
{\color[HTML]{000000} \textbf{No.}} & {\color[HTML]{000000} \textbf{\begin{tabular}[c]{@{}c@{}}Time\\ (mins)\end{tabular}}} & {\color[HTML]{000000} \textbf{\begin{tabular}[c]{@{}c@{}}T-table\\ index\end{tabular}}} & {\color[HTML]{000000} \textbf{\begin{tabular}[c]{@{}c@{}}Value\\ before flip\end{tabular}}} & {\color[HTML]{000000} \textbf{\begin{tabular}[c]{@{}c@{}}Value\\ after flip\end{tabular}}} \\ \hline
{\color[HTML]{000000} 1} & {\color[HTML]{000000} 1035} & {\color[HTML]{000000} Te1[139]} & {\color[HTML]{000000} 477a 3d3d} & {\color[HTML]{000000} 47fa 3d3d} \\ \hline
{\color[HTML]{000000} 2} & {\color[HTML]{000000} 538} & {\color[HTML]{000000} Te0[25]} & {\color[HTML]{000000} b3d4 d467} & {\color[HTML]{000000} a3d4 d467} \\ \hline
{\color[HTML]{000000} 3} & {\color[HTML]{000000} 224} & {\color[HTML]{000000} Te1[254]} & {\color[HTML]{000000} d66d bbbb} & {\color[HTML]{000000} c66d bbbb} \\ \hline
{\color[HTML]{000000} 4} & {\color[HTML]{000000} 3623} & {\color[HTML]{000000} Te1[38]} & {\color[HTML]{000000} ae3d 9393} & {\color[HTML]{000000} ae3d 9193} \\ \hline
{\color[HTML]{000000} 5} & {\color[HTML]{000000} 12} & {\color[HTML]{000000} Te3[87]} & {\color[HTML]{000000} cbcb 468d} & {\color[HTML]{000000} cbcb 460d} \\ \hline
{\color[HTML]{000000} 6} & {\color[HTML]{000000} 105} & {\color[HTML]{000000} Te3[148]} & {\color[HTML]{000000} 2222 6644} & {\color[HTML]{000000} 0222 6644} \\ \hline
{\color[HTML]{000000} 7} & {\color[HTML]{000000} 256} & {\color[HTML]{000000} Te1[88]} & {\color[HTML]{000000} bed4 6a6a} & {\color[HTML]{000000} bed4 2a6a} \\ \hline
{\color[HTML]{000000} 8} & {\color[HTML]{000000} 67} & {\color[HTML]{000000} Te1[193]} & {\color[HTML]{000000} 88f0 7878} & {\color[HTML]{000000} 88f0 7868} \\ \hline
\end{tabular}
}
\caption{Faults induced by ExplFrame on AES T-tables}
\label{tab:results}
\vspace{-0.1cm}
\end{table}
%
%
In order to elaborate the relation of the attack with the number of ciphertexts, we present a convergence plot in Fig.~\ref{fig:plot_sei_convergence}.
%
The blue region in this plot presents the SEI values for wrong key guesses. The red line, on the other hand, refers to the SEI value corresponding to the correct key. For the attack to be successful, the red line should remain over the blue region. One may observe that in our experiment, the red line crosses the blue region near $7000$ ciphertexts and the separation becomes prominent with $8000$ ciphertexts. This clearly establishes the efficacy of the DRPFA. 
\section{Discussion}
\label{sec: discussion}
The attack use-case presented in the last section assumes that the AES is implemented by means of T-tables. Certain modern AES implementations use dedicated hardware accelerators, such as AES-NI~\cite{aesni} instead of T-tables. However, this fact does not rule out the efficacy of the proposed attack strategy as it is not specific to AES. In contrast, the hardware accelerators available in modern processors provide support for AES only. Moreover, there exist cryptographic primitives such as pseudorandom number generators which still utilize the T-tables in OpenSSL implementation~\cite{cohney_prn}. The proposed attack thus remains completely relevant for modern systems.     

Error Correcting Codes (ECC) are often considered as effective countermeasures against Rowhammer faults~\cite{kim_isca14}. However, we claim that a straightforward application of ECC as in~\cite{kim_isca14} cannot prevent information leakage by means of faults. This claim is based on the observation~\cite{cojocar_ecc} that the error correction operation takes considerably large number of clock cycles with respect to the normal operation~\cite{kwong_rambleed}. It is thus feasible to detect and pinpoint when error correction takes place during the encryption (with a timing channel) by using some of the strategies described in~\cite{cachegames, flush_reload}. Interestingly, if the attacker possess the knowledge of the exact corrupted entry within the faulted T-table~{\footnote{which is reasonable if the T-table is kept in a shared memory, and attacker have a read access to it.}}, she can figure out the input to the table. Figuring out the table input is equivalent to the key recovery as the secret intermediate states of a block cipher get exposed by this mean. Moreover, this \emph{timing-assisted fault analysis} strategy does not require any access to the ciphertexts of the encryption block which enhances its scope for certain other classes of crypto-primitives. Future work will present a practical realization of this attack on state-of-the-art systems.

%% file: source/conclusion.tex
\section{Conclusion}
\label{sec:conclusion}
In this paper, we presented ExplFrame, a novel attack technique that combines the vulnerability of page frame cache with Rowhammer to induce faults in victim process's data, entirely from user space. We highlighted how PFC can be used as an attack vector to restrict a process to attacker-chosen locations in DRAM. To validate the practicality of our claims, we demonstrated an end-to-end attack on OpenSSL AES to induce faults in T-tables. We also presented an improvised Fault Analysis technique to extract the key from the faulty ciphertexts, created due to faulted T-tables.
